\documentclass[12pt]{article}
\usepackage{amssymb,dsfont,amsthm,amsmath,graphics,colortbl,epsfig,eepic}

\def\bbbc{{\Bbb C}}

\def\bbbz{{\Bbb Z}}

\def\ad{\mbox{ad}\,}
\def\tr{\mbox{tr}\,}

\def\Ad{\mbox{Ad}\,}
\def\Aut{\mbox{Aut}\,}

\def\openone{\leavevmode\hbox{\small1\kern-3.3pt\normalsize1}}

\def\newpic#1{%
   \def\emline##1##2##3##4##5##6{%
      \put(##1,##2){\special{em:point #1##3}}%
      \put(##4,##5){\special{em:point #1##6}}%
      \special{em:line #1##3,#1##6}}}

\newpic{}

\begin{document}
\arraycolsep=2pt

\numberwithin{equation}{section}
\allowdisplaybreaks
\bibliographystyle{plain}

\begin{center}
{\large \bf The Generalised Zakharov-Shabat System and the Gauge Group Action}

{\bf  Georgi G. Grahovski$^{1,2}$}

\medskip

{\it $^{1}$ School of Mathematical Sciences, Dublin Institute of Technology, \\
Kevin Street, Dublin 8, Ireland} \\
{\it $^2$ Institute of Nuclear Research and Nuclear Energy,  Bulgarian Academy of Sciences,\\ 72 Tsarigradsko chausee, Sofia 1784, Bulgaria } \\

\medskip

{\small E-mail: {\tt georgi.grahovski@dit.ie} $\quad$ {\tt grah@inrne.bas.bg}}

\end{center}

\begin{abstract}
The generalized Zakharov--Shabat systems with
complex-valued {\it non-regular} Cartan elements and the systems
studied by Caudrey, Beals and Coifman (CBC systems) and their
gauge equivalent are studied. This study includes: the properties
of fundamental analytical solutions (FAS) for the gauge-equivalent
to CBC systems and the minimal set of scattering data; the
description of the class of nonlinear evolutionary equations,
solvable by the inverse scattering method, and the recursion
operator, related to such systems; the hierarchies of Hamiltonian
structures. The results are illustrated on the example of the
multi-component nonlinear Schr\"odinger (MNLS) equations and the
corresponding gauge-equivalent multi-component Heisenberg
ferromagnetic (MHF) type models, related to $so(5,\bbbc)$ algebra.

\end{abstract}

\section{Introduction}\label{sec:1}

The multi-component Zakharov-Shabat (ZS) system leads to such
important systems as the multi-com\-po\-nent non-linear
Schr\"odinger equation (NLS), the $N$-wave type equations, etc.
All of these systems are integrable via the inverse scattering
method.  In the class of nonlinear evolution equations (NLEE)
related to the Zakharov--Shabat (ZS) system \cite{Za*Sh,1}, the
Lax operator belonging to $sl(2,\bbbc) $ algebra is studied. This
class of NLEE contains  physically important equations as the
nonlinear Schr\"odinger equation (NLS), the sine-Gordon,
Korteweg--de-Vriez (KdV) and the modified Korteweg--de-Vriez
(mKdV) equations. In the recent  years, the gauge equivalent systems to various versions of the generalised Zakharov-Shabat have been systematically studied \cite{Ferap,vg1,vg2,vgn,vgn1,ours,ours2,ours3,ours4,ours5,yan1,yan2,yan3,wu}. Recently, it was also shown \cite{cil2010} that the spectral problem for the Degasperis-Procesi equation can be cast into Zakharov-Shabat form with an $sl(3, \bbbc)$ Lax pair  with additional $\bbbz_3$ and $\bbbz_2$ symmetries.

Here, we consider the $n \times n $ system
\cite{BS,Caud*80,ForGib*80b}:
\begin{eqnarray}\label{eq:1.5} L\Psi
(x,t,\lambda )= \left({\rm i}{{\rm d} \over {\rm d}
x}+Q(x,t)-\lambda J\right) \Psi (x,t,\lambda ),
\end{eqnarray} where the potential $Q(x,t) $  takes values in the
semi-simple Lie algebra ${\frak g} $
\cite{MiOlPer*81,G*86,Za*Mi,ForKu*83}:
\begin{eqnarray}\label{eq:1.6}
Q(x,t) = \sum_{\alpha \in \Delta _+} \left(q_{\alpha
}(x,t)E_{\alpha }+ q_{-\alpha }(x,t)E_{-\alpha } \right)\in {\frak
g}_J \qquad J = \sum_{j=1}^{r}a_jH_j \in {\frak h}. \nonumber
\end{eqnarray}
For the case of complex $J$, we refer to this system as the
Caudrey-Beals-Coifman (CBC) system. Here, $J $ is a {\it non-regular}
element in the Cartan subalgebra ${\frak h} $ of ${\frak g} $,
${\frak g}_J $ is the image of $\mbox{ad}_J $, $\{E_{\alpha },
H_i\} $ form the Cartan--Weyl Basis in ${\frak g} $, $\Delta _+ $
is the set of positive roots of the algebra, $r= \mbox{rank}\,
{\frak g}= \mbox{dim}\, {\frak h}$. For more details, see section
2 below. The {\it non-regularity} of the Cartan elements means
that ${\frak g}_J $ is {\it not} spanned by all root vectors
$E_{\alpha } $ of ${\frak g} $, i.e. $\alpha (J) \neq 0 $ for any
root $\alpha $ of ${\frak g} $.

The given NLEE, as well as the other members of its hierarchy, possess
a Lax representation of the form (according to (\ref{eq:1.5})):
$
[L(\lambda ), M_P(\lambda )]=0,
$
where
\begin{eqnarray}\label{eq:1.7}
M_P\Psi (x,t,\lambda ) = \left({\rm i}{{\rm d} \over {\rm d} t}+\sum_{k=-S}^{P-1}
V_k(x,t)-\lambda ^Pf_PI \right) \Psi (x,t,\lambda )=0, \qquad
I\in {\frak h},
\end{eqnarray}
which must hold identically with respect to $\lambda  $. A
standard procedure  generalising the one proposed by Ablowitz, Kaup, Newell and Segur (AKNS)  \cite{AKNS} allows
us to evaluate $V_k(x,t) $ in terms of $q(x,t) $ and its
$x$-derivatives. Here and below, we consider only the class of
potentials $q(x,t) $ vanishing fast enough for $|x|\rightarrow
\infty  $. Then, one may also check that the asymptotic value of
the potential in $M_P(\lambda )$, namely $f^{(P)}(\lambda
)=f_P\lambda ^PI$, may be understood as the dispersion law of the
corresponding NLEE.

Another important step  in the development of the Inverse Scattering Method (ISM) is the introduction of
the reduction group by A. V. Mikhailov \cite{2},
and further developed in
\cite{ForGib*80b,ForKu*83,Za*Mi,MiOlPer*81}. This allows one to
prove that some of the well known models in  field theory
\cite{2}, and also a number of new interesting NLEE
\cite{2,ForGib*80b,MiOlPer*81,GEI}, are integrable by the ISM and
possess special symmetry properties. As a result, its potential
$q(x,t) $ has a very special form and $J$ {\it can no-longer be
chosen real}.

This problem of constructing the spectral theory for
(\ref{eq:1.5}) in the most general case, when $J $ has an
arbitrary complex eigenvalues, was initialized by Beals, Coifman
and Caudrey \cite{BC,BC2,BC3,Caud*80}, and continued by Zhou
\cite{Zhou} in the case when the algebra ${\frak g} $ is $ sl(n)
$, $q(x,t) $ vanishing fast enough for $|x| \rightarrow \infty$,
and no {\it a priori} symmetry conditions are imposed on $q(x,t)
$. This was done later for any semi-simple Lie algebras by
Gerdjikov and Yanovski \cite{VYa}.

The applications of the differential geometric and Lie algebraic
methods to soliton type equations lead to the discovery of a close
relationship between the multi-component (matrix) NLS equations
and the symmetric and homogeneous spaces \cite{ForKu*83}. In
\cite{ForKu*83} it was shown that the integrable MNLS systems have
a Lax representation of the form (\ref{eq:1.5}), where $J $ is a
constant element of the Cartan subalgebra $\mathfrak{h} \subset
\mathfrak{g} $ of the simple Lie algebra $\mathfrak{g} $ and
$Q(x,t) \equiv [J,\widetilde{Q}(x,t)] \in
\mathfrak{g}/\mathfrak{h}$. In other words, $Q(x,t) $ belongs to
the co-adjoint orbit $\mathcal{M}_J $ of $\mathfrak{g} $ passing
through $J$. Later on, this approach was extended to other types
of multi-component integrable models, like the derivative NLS,
Korteweg-de-Vriez and modified Korteweg-de-Vriez, $N$-wave,
Davey-Stewartson, Kadomtsev-Petviashvili equations
\cite{AthFor,For}.

The choice of $J $ determines the dimension of $\mathcal{M}_J $
which can be viewed as the phase space of the relevant nonlinear
evolution equations (NLEE).  It is is equal to the number of roots
of $ \mathfrak{g} $ such that $\alpha (J)\neq 0 $.  Taking into
account that if $\alpha  $ is a root, then  $-\alpha $ is also a
root of $\mathfrak{g} $, then $\dim \mathcal{M}_J $ is always
even \cite{loos}.

The degeneracy of $J $ means that the subalgebra $\mathfrak{g}_J
\subset  \mathfrak{g} $ of elements commuting with $J $ (i.e., the
kernel of the  operator $\ad_J $) is non-commutative.  This makes
more difficult the derivation of the fundamental analytic
solutions (FAS) of the Lax operator (\ref{eq:1.5}) and the
construction of the corresponding (generating) recursion operator
$\Lambda$. The explicit construction of the recursion operator
related to (\ref{eq:1.5}), using the gauge covariant approach
\cite{G*86,VSG*94}, is outlined in\cite{vgn2}.

As we mentioned above the Lax operator for the MNLS equations
formally has the form (\ref{eq:1.5}) but now $J $ is no longer a
regular element of $\mathfrak{h} $. This means that the subalgebra
$\mathfrak{g}_J\subset \mathfrak{g} $ of elements commuting with
$J $ (i.e., the kernel of the operator $\ad_J $) is a
non-commutative one. The dispersion law of the MNLS eqn. is
quadratic in $\lambda  $: $f_{\rm MNLS}=2\lambda ^2 J $. The
general form of the MNLS equations and their $M $-operators is:
\begin{eqnarray}\label{eq:MNLS}
&& i {dq \over dt } + 2\ad_J^{-1}{d^2q  \over dx^2 }  +[q,
\pi_0[q, \ad_J^{-1}q]] -2i(\openone - \pi _0)[q, \ad_J^{-1}q_x]
=0, \\\label{eq:M-NLS} && M(\lambda)\psi \equiv \left(i{d\over dt}
- V_0^{\rm d} + 2i \mbox{ad}_J^{-1}q_x(x,t) +2\lambda q(x,t)
-2\lambda ^2 J \right) \psi (x,t,\lambda )=0,
\end{eqnarray}
where $V_0^d=\pi_0 \left( [q, \ad_J^{-1}q_x]\right)$ and $\pi_0$
is the projector onto $\mathfrak{g}_J $; (see also Section 4.1
below).

The zero-curvature condition $ [L(\lambda ), M_P(\lambda )]=0, $
is invariant under the action of the group of gauge
transformations \cite{ZaTa}. Therefore, the gauge equivalent
systems are again completely integrable, possess a hierarchy of
Hamiltonian structures, etc, \cite{FaTa,1,VYa,ZaTa}.

The structure of this paper is as follows: In Section 2 we
summarize some basic facts about the reduction group and Lie
algebraic details. The construction of the fundamental analytic
solutions (FAS) is sketched in Section 3 which is done separately
for the case of real Cartan elements (Section 3.1) and for complex
ones (Section 3.2). The gauge equivalent MHF's to the MNLS systems
are described in Section 4. In Section 5 we present an example of
a MNLS type system, related to the $so(5)$ Lie algebra and its
corresponding MHF one.

The present article presents an extension of our results
\cite{gc}.

\section{Preliminaries}\label{sec:2}

\subsection{Simple Lie Algebras}\label{ssec:2.1}

Here, we fix up the notations and the normalization conditions for
the Cartan-Weyl generators of ${\frak g} $ \cite{Helg}. We
introduce $h_k\in {\frak h} $, $k=1,\dots,r $ and $E_\alpha  $,
$\alpha \in \Delta$, where $\{h_k\} $ are the Cartan elements dual
to the orthonormal basis $\{e_k\}$ in the root space ${\mathbb
E}^r $. Along with $h_k $, we introduce also
\begin{equation}\label{eq:31.1}
H_\alpha = {2  \over (\alpha ,\alpha ) } \sum_{k=1}^{r} (\alpha ,e_k) h_k,
\quad \alpha \in \Delta ,
\end{equation}
where $(\alpha ,e_k) $ is the scalar product in the root space
${\mathbb E}^r $ between the root $\alpha  $ and $e_k $. The commutation
relations are given by \cite{LA}:
\begin{eqnarray*}\label{eq:31.2} [h_k,E_\alpha
] = (\alpha ,e_k) E_\alpha , \quad [E_\alpha ,E_{-\alpha }]=H_\alpha ,
\quad [E_\alpha ,E_\beta ] = \left\{ \begin{array}{ll} N_{\alpha
,\beta } E_{\alpha +\beta } \quad & \mbox{for}\; \alpha +\beta \in \Delta
\\ 0 & \mbox{for}\; \alpha +\beta \not\in \Delta \cup\{0\}. \end{array}
\right.  \end{eqnarray*}

We will denote by $\vec{a}=\sum_{k=1}^{r}a_k e_k $ the $r $-dimensional
vector dual to $J\in {\frak h} $; obviously $J=\sum_{k=1}^{r}a_k h_k $. If $
J  $ is a regular real element in ${\frak h}$, then without restrictions we
may  use it to introduce an ordering in $\Delta  $. Namely, we will
say that the root $\alpha \in\Delta _+ $ is positive (negative) if
$(\alpha ,\vec{a})>0 $ ($(\alpha ,\vec{a})<0 $ respectively).
The normalization of the basis is determined by:
\begin{eqnarray}\label{eq:32.1}
E_{-\alpha } =E_\alpha ^T, \quad \langle E_{-\alpha },E_\alpha \rangle
={2  \over (\alpha ,\alpha ) }, \quad
N_{-\alpha ,-\beta } = -N_{\alpha ,\beta }, \quad N_{\alpha ,\beta } =
\pm (p+1),
\end{eqnarray}
where the integer $p\geq 0 $ is such that $\alpha +s\beta
\in\Delta  $ for all $s=1,\dots,p $ $ \alpha +(p+1)\beta
\not\in\Delta  $ and $\langle \cdot,\cdot \rangle  $ is the
Killing form of ${\frak  g}$ \cite{GoGr,Helg}.  The root system
$\Delta $ of ${\frak g} $ is invariant with respect to the Weyl
reflections $A^*_\alpha $; on the vectors $\vec{y}\in {\mathbb
E}^r $ they act as $ A^*_\alpha \vec{y} = \vec{y} - {2(\alpha
,\vec{y})  \over (\alpha ,\alpha )} \alpha , \quad \alpha \in
\Delta$. All Weyl reflections $A^*_\alpha $ form a finite group
$W_{{\frak g}} $ known as the Weyl group. One may introduce, in a
natural way, an action of the Weyl group on the Cartan-Weyl basis,
namely \cite{FG,Hu}:
\begin{eqnarray*}\label{eq:32.3}
A^*_\alpha (H_\beta ) \equiv A_\alpha
H_\beta A^{-1}_{\alpha } = H_{A^*_\alpha \beta }, \qquad A^*_\alpha
(E_\beta ) \equiv A_\alpha E_\beta A^{-1}_{\alpha } = n_{\alpha ,\beta }
E_{A^*_\alpha \beta }, \quad n_{\alpha ,\beta }=\pm 1.
\end{eqnarray*}
It is also well known that the matrices $A_\alpha  $ are given (up to a
factor from the Cartan subgroup) by
$
A_\alpha ={\rm e}^{E_\alpha } {\rm e}^{-E_{-\alpha }} {\rm e}^{E_\alpha } H_A,
$
where $H_A $ is a conveniently chosen element from the Cartan
subgroup such that $H_A^2=\openone  $.

As we already mentioned in the Introduction, the MNLS equations
correspond to the Lax operator (\ref{eq:1.5}) with non-regular
(constant) Cartan elements $J\in \mathfrak{h}$.  If $J$ is a
regular element of the Cartan subalgebra of $\mathfrak{g}$, then
$\ad_J$ has as many different eigenvalues as is the number of the
roots of the algebra and they are given by $a_j=\alpha_j(J)$,
$\alpha_j\in \Delta$. Such $J $'s can be used to introduce
ordering in the root system by assuming that $\alpha >0 $ if
$\alpha (J)>0 $. In what follows, we will assume that all roots
for which $\alpha (J)>0 $ are positive.

Obviously, we can consider the eigensubspaces of $\ad_J $ as a
grading of the algebra $\mathfrak{g} $. In what follows, we will
consider symmetric spaces related to maximally degenerated $J $,
i.e. $\ad_J $ has only four non-vanishing eigenvalues: $ \pm a $
and $ \pm 2a $. Then $\mathfrak{g}$ is split into a direct sum of
the subalgebra $\mathfrak{g}_0 $ and the linear subspaces
$\mathfrak{g}_{\pm } $:
\begin{eqnarray}\label{eq:KF14.1}
\mathfrak{ g} &=& \mathfrak{ g}_0  \oplus \mathfrak{ g}_+ \oplus
\mathfrak{g}_{-},\qquad \mathfrak{ g}_{\pm } =
\mathop{\oplus}\limits_{k=1}^2 \mathfrak{ g}_{\pm k}  \qquad
\mathfrak{ g}_{\pm } =  \mbox{l.c.} \left\{ X_{\pm j}\,|\, [J,
X_{\pm }]=\pm ka X_{\pm } \right\}, \qquad k=1,2. \nonumber
\end{eqnarray}
The subalgebra  $\mathfrak{ g}_0 $  contains the Cartan subalgebra
$\mathfrak{h} $ and also all root vectors $E_{\pm \alpha }\in
\mathfrak{g} $ corresponding to the roots $\alpha  $ such that
$\alpha (J)=(\vec{a},\alpha )=0 $. The root system $\Delta  $ is
split into subsets of roots $\Delta = \theta _0\cup \theta _+\cup
(-\theta _+)$, where:
\begin{eqnarray}\label{eq:KF14.3}
\theta _0 &=& \left\{ \alpha \in \Delta \,|\, \alpha (J) =0
\right\}, \quad \theta _+ = \left\{ \alpha \in \Delta \,|\, \alpha
(J)  >0 \right\}.
\end{eqnarray}

\subsection{The Reduction Group}\label{ssec:2.2}

The principal idea underlying Mikhailov's reduction group \cite{2}
is to impose algebraic restrictions on the Lax operators $L $ and
$M$, which will be automatically compatible with the corresponding
equations of motion. Due to the purely Lie-algebraic nature of the
Lax representation, this is most naturally done by embedding the
reduction group as a subgroup of $\mbox{Aut}\, {\frak g} $ -- the
group of automorphisms of ${\frak g} $. Obviously, to each
reduction imposed on $L $ and $M$, there will correspond a
reduction of the space of fundamental solutions ${\bf S}_\Psi
\equiv \{\Psi (x,t,\lambda )\} $ of (\ref{eq:1.5}).

Some of the simplest ${\mathbb Z}_2 $-reductions of Zakharov--Shabat
systems have been known for a long time (see \cite{2}) and are related to
outer automorphisms of ${\frak g} $ and ${\frak G} $, namely:
\begin{eqnarray}\label{eq:C-1}
C_1\left( \Psi (x,t,\lambda ) \right) =  A_1 \Psi^\dag (x,t,\kappa
(\lambda )) A_1^{-1} = \widetilde{\Psi}^{-1}(x,t,\lambda ), \qquad
\kappa (\lambda )=\pm \lambda ^*,
\\
C_2\left( \Psi (x,t,\lambda ) \right) =  A_3 \Psi^* (x,t,\kappa
(\lambda )) A_3^{-1} = \widetilde{\Psi}(x,t,\lambda ),
\end{eqnarray}
where $A_1$ and $A_3$ are elements of the group of authomorphisms
$\mbox{Aut}\,{\frak g}$ of the algebra ${\frak g}$. Since our aim
is to preserve the form of the Lax pair, we limit ourselves to
automorphisms preserving the Cartan subalgebra ${\frak  h} $. The
reduction group, $G_R$, is a finite group which preserves the Lax
representation, i.e. it ensures that the reduction constraints are
automatically compatible with the evolution. $G_R $ must have two
realizations:

i) $G_R \subset {\rm Aut}{\frak g} $

ii) $G_R
\subset {\rm Conf}\, \Bbb C $, i.e. as conformal mappings of the complex
$\lambda $-plane.

To each $g_k\in G_R $, we relate a reduction
condition for the Lax pair as follows \cite{2}:
\begin{equation}\label{eq:2.1}
C_k(U(\Gamma _k(\lambda ))) = \eta_k U(\lambda ),
\end{equation}
where $U(x,\lambda )=q(x)-\lambda J $, $C_k\in \mbox{Aut}\; {\frak g} $ and
$\Gamma _k(\lambda )$ are the images of $g_k $ and $\eta_k =1 $ or $-1 $
depending on the choice of $C_k $.  Since $G_R $ is a finite group, then
for each $g_k$, there exists an integer $N_k $ such that $g_k^{N_k}
=\openone $.

It is well known that $\Aut {\frak g} \equiv V\otimes \Aut_0 {\frak g}$, where
$V $ is the group of outer automorphisms (the symmetry group of the Dynkin
diagram) and $\Aut_0 {\frak g} $ is the group of inner automorphisms. Since
we start with $I,J\in {\frak h}$, it is natural to consider only those inner
automorphisms that preserve the Cartan subalgebra ${\frak h} $. Then $\Aut_0
{\frak g} \simeq \Ad_H \otimes W $ where $\Ad_H $ is the group of similarity
transformations with elements from the Cartan subgroup
and $W $ is the Weyl group of ${\frak g} $.

Generically, each element $g_k\in G $ maps $\lambda  $ into a
fraction-linear function of $\lambda  $. Such action however is
appropriate for a more general class of Lax operators which are fraction
linear functions of $\lambda  $.

\section{The Caudrey--Beals--Coifman systems}\label{sec:3}

\subsection{Fundamental analytical solutions and scattering data for real $J
$.}
\label{3.1}

The direct scattering problem for the Lax operator (\ref{eq:1.5}) is based
on the Jost solutions:
\begin{eqnarray}\label{eq:3.1.1}
\lim_{x \to \infty }\psi (x,\lambda ){\rm e}^{{\rm i}\lambda Jx}=\openone , \qquad
\lim_{x \to -\infty }\phi (x,\lambda ){\rm e}^{{\rm i}\lambda Jx}=\openone ,
\end{eqnarray}
and the scattering matrix
\begin{eqnarray}\label{eq:3.1.2}
T(\lambda )=(\psi (x,\lambda ))^{-1}\phi (x,\lambda ).
\end{eqnarray}
The fundamental analytic solutions (FAS) $\chi^{\pm} (x,\lambda ) $ of
$L(\lambda ) $ are analytic functions of $\lambda  $ for
$\mbox{Im}\,\lambda \gtrless 0$ and are related to the Jost solutions by
\cite{G*86,GVY*08}
\begin{eqnarray}\label{eq:3.1.3}
\chi ^{\pm}(x,\lambda )=\phi (x,\lambda )S_J^{\pm}(\lambda )= \psi
^{\pm}(x,\lambda )T_J^{\mp}(\lambda )D_J^{\pm}(\lambda ),
\end{eqnarray}
where $T_J^{\pm}(\lambda ) $, $S_J^{\pm}(\lambda ) $ and
$D_J^{\pm}(\lambda ) $ are the factors of the Gauss decomposition
of the scattering matrix:
\begin{eqnarray}\label{eq:3.1.4}
&&T(\lambda )=T_J^-(\lambda )D_J^+(\lambda )\hat{S}_J^{+}(\lambda
)=
T_J^+(\lambda )D_J^-(\lambda )\hat{S}_J^-(\lambda ) \\
&&T_J^{\pm}(\lambda )=\exp \left(\sum_{\alpha >0}t^{\pm}_{\pm
\alpha ,J}(\lambda )E_{\alpha } \right), \qquad S_J^{\pm}(\lambda
)=\exp \left(\sum_{\alpha >0}s^{\pm}_{\pm \alpha,J
}(\lambda )E_{\alpha } \right),  \nonumber\\
&&D_J^{+}(\lambda )=I\exp \left(\sum_{j=1}^{r}{2d_J^+(\lambda )
\over (\alpha _j,\alpha _j)}H_j \right), \qquad D_J^{-}(\lambda
)=I\exp \left(\sum_{j=1}^{r}{2d_J^-(\lambda )  \over (\alpha
_j,\alpha _j)}H_j^- \right). \nonumber
\end{eqnarray}
Here, $H_j=H_{\alpha _j} $, $H_j^-=w_0(H_j) $, $\hat{S}\equiv
S^{-1} $, $I$ is an element from the universal center of the
corresponding Lie group ${\frak  G} $ and the superscript $+ $ (or
$- $) in the Gauss factors means upper- (or lower-)
block-triangularity for $T_J^{\pm} (\lambda )  $, $S_J^{\pm}
(\lambda ) $  and shows that $D_J^{+} (\lambda )  $ (or $D_J^{-}
(\lambda ) $) are analytic functions with respect to $\lambda  $
for $\mbox{Im}\, \lambda  >0 $ (or $\mbox{Im}\, \lambda <0$,
respectively).

On the real axis $\chi ^+(x,\lambda )$ and $\chi ^-(x,\lambda ) $
are linearly related by:
\begin{eqnarray}\label{eq:3.1.6}
\chi ^+(x,\lambda )=\chi ^-(x,\lambda )G_{J,0}(\lambda ), \qquad
G_{J,0}(\lambda )=S_J^+(\lambda )\hat{S}_J^-(\lambda ),
\end{eqnarray}
and the sewing function $G_{J,0}(\lambda ) $ may be considered as
a minimal system of scattering data provided that the Lax operator
(\ref{eq:1.5}) has no discrete eigenvalues \cite{G*86}.

\subsection{The CBC Construction for Semisimple Lie Algebras}
\label{3.2}

Here, we will sketch the construction of the FAS for the case of
complex-valued regular Cartan elements $J$: $\alpha (\psi )\neq 0$,
following the general ideas of Beals and Coifmal \cite{BC,BDT} for
the $sl(n) $ algebras and \cite{VYa} for the orthogonal and
symplectic algebras. These ideas consist of the following:
\begin{enumerate}

\item For potentials $q(x) $ with small norm $||q(x)||_{L^1} <1$, one can
divide the complex $\lambda  $--plane into sectors and then construct an
unique FAS $m_{\nu }(x,\lambda ) $ which is analytic in each of these
sectors $\Omega _{\nu } $;

\item For these FAS in each sector, there is a certain Gauss decomposition
problem for the scattering matrix $T(\lambda ) $ which has a unique
solution in the case of absence of discrete eigenvalues.

\end{enumerate}
The main difference between the cases of real-valued and complex-valued $
J $ lies in the fact that for complex $J $ the Jost solutions and the
scattering data exist only for the potentials on compact support.

We define the regions (sectors) $\Omega _{\nu } $ as consisting of
those $ \lambda  $'s for which $\mbox{Im}\, (\lambda \alpha
(J))\neq 0 $ for any $ \alpha \in \Delta  $. Thus, the boundaries
of the $\Omega _{\nu } $'s consist of the set of straight lines:
\begin{eqnarray}\label{eq:3.2.1}
l_{\alpha }\equiv \{ \lambda : \mbox{Im}\, \lambda \alpha (J)=0, \qquad
\alpha \in \Delta \},
\end{eqnarray}
and to each root $\alpha$, we can associate a certain line
$l_{\alpha } $; different roots may define coinciding lines.

Note that with the change from $\lambda  $ to $\lambda {\rm e}^{{\rm i}\eta } $
and $ J $ to $J{\rm e}^{-{\rm i}\eta } $ (this leads the product $\lambda \alpha
(J) $ invariant), we can always choose $l_1 $ to be along the positive real
$\lambda  $ axis.

To introduce an ordering in each sector $\Omega _{\nu }$, we choose the
vector $\vec{a}_{\nu }(\lambda )\in {\mathbb E}^r $ to be dual to the
element
$\mbox{Im}\, \lambda J \in {\frak  h} $. Then, in each sector we split
$\Delta  $ into
\begin{eqnarray}\label{eq:3.2.2}
\Delta =\Delta _{\nu }^+ \cup \Delta _{\nu }^-, \qquad
\Delta _{\nu }^{\pm}=\{\alpha \in \Delta : \mbox{Im}\, \lambda \alpha
(J)\gtrless 0,\, \lambda \in \Omega _{\nu }\}.
\end{eqnarray}
If $\lambda \in \Omega _{\nu } $ then $-\lambda \in \Omega _{M+\nu
} $ (if the lines $l_{\alpha } $ split the complex $\lambda
$-plane into $2M $ sectors). We also need  the subset of roots:
\begin{eqnarray}\label{eq:3.2.2b}
\delta _{\nu }=\{\alpha \in \Delta \, : \, \mbox{Im}\, \lambda \alpha
(J)=0, \, \lambda \in l_{\nu }\}
\end{eqnarray}
which will be a root system of some subalgebra ${\frak  g}_{\nu }\subset
{\frak  g} $. Then, we can write that
\begin{eqnarray}\label{eq:3.2.3}
{\frak  g}= \mathop{\oplus}\limits_{\nu =1}^{M} {\frak  g}_{\nu }  \qquad
\Delta = \mathop{\cup}\limits_{\nu =1}^{M}\delta _\nu  \qquad
\delta _{\nu }=\delta _{\nu }^+ \cup \delta _{\nu }^-, \qquad
\delta _{\nu }^{\pm}=\delta _{\nu }\cap \Delta _{\nu }^\pm \nonumber .
\end{eqnarray}
Thus, we can describe in more detail the sets $\Delta _{\nu }^{\pm} $:
\begin{eqnarray}\label{eq:3.2.4}
\Delta _k^+=\delta _1^+ \cup \delta _2^+ \cup \dots \cup \delta _k^+
\cup \delta _{k+1}^-\cup \dots \cup \delta _M^-, \quad \Delta _{k+M}^+=\Delta _k^-, \quad
k=1,\dots , M.
\end{eqnarray}
Note that each ordering in $\Delta  $ can be obtained from the
"canonical" one by an action of a properly chosen element of the
weyl group ${\frak W}({\frak  g}) $.

Now, in each sector $\Omega _{\nu }$, we introduce the FAS $\chi _{\nu
}(x,\lambda ) $ and $m_{\nu }(x,\lambda )=\chi _{\nu }(x,\lambda
){\rm e}^{{\rm i}\lambda Jx} $ satisfying the equivalent equation:
\begin{eqnarray}\label{eq:3.2.5}
{\rm i}{{\rm d} m_{\nu }  \over {\rm d} x} + q(x)m_{\nu }(x,\lambda )-\lambda
[J,m_{\nu }(x,\lambda )]=0, \qquad  \lambda \in \Omega _{\nu }.
\end{eqnarray}
If $q(x) $ is a potential on compact support, then the FAS $m_{\nu
}(x,\lambda ) $ are related to the Jost solutions by
\begin{eqnarray}\label{eq:3.2.6}
&&m_{\nu }(x,\lambda )=\phi (x,\lambda )S_{J,\nu }^+(\lambda ){\rm e}^{{\rm i}
\lambda Jx} =\psi (x,\lambda )T_{J,\nu }^-(x,\lambda )D_{J,\nu }^+(\lambda
){\rm e}^{{\rm i}\lambda Jx}, \\
&&m_{\nu -1}(x,\lambda )=\phi (x,\lambda )S_{J,\nu
}^-(\lambda ){\rm e}^{{\rm i}\lambda Jx} =\psi (x,\lambda )T_{J,\nu }^+(x,\lambda
)D_{J,\nu }^-(\lambda ){\rm e}^{{\rm i}\lambda Jx}, \qquad \lambda \in l_{\nu }.
\nonumber
\end{eqnarray}
From the definitions of $m_{\nu }(x,\lambda ) $
and the scattering matrix $T(\lambda )$, we have
\begin{eqnarray}\label{eq:3.2.7}
T(\lambda )=T_{J,\nu }^-(\lambda )D_{J,\nu }^+(\lambda )\hat{S}_{J,\nu }^+
(\lambda )= T_{J,\nu }^+(\lambda )D_{J,\nu }^-(\lambda )\hat{S}_{J,\nu }^-
(\lambda ) , \quad \lambda \in l_{\nu }
\end{eqnarray}
where, in the first equality, we take $\lambda =\mu {\rm e}^{{\rm
i}0} $ and for the second-- $\lambda =\mu {\rm e}^{-{\rm i}0} $
with $\mu \in l_{\nu } $. The corresponding expressions for the
Gauss factors have the form:
\begin{eqnarray}\label{eq:3.2.8}
&&S_{J,\nu }^+(\lambda )=\exp \left(\sum_{\alpha \in \Delta _{\nu
}^+}s_{\nu ,\alpha }^+(\lambda )E_{\alpha } \right), \qquad
S_{J,\nu }^-(\lambda )=\exp \left(\sum_{\alpha \in \Delta _{\nu
-1}^+}s_{\nu ,\alpha }^-(\lambda )E_{-\alpha } \right),  \nonumber\\
&&T_{J,\nu }^+(\lambda )=\exp \left(\sum_{\alpha \in \Delta _{\nu
-1}^+}t_{\nu ,\alpha }^+(\lambda )E_{\alpha } \right), \qquad
T_{J,\nu }^-(\lambda )=\exp \left(\sum_{\alpha \in \Delta _{\nu
}^+}t_{\nu ,\alpha }^-(\lambda )E_{-\alpha } \right),  \nonumber\\
&&D_{J,\nu }^+(\lambda )=\exp ({\bf d}_{\nu }^+(\lambda )\cdot {\bf
H}_{\nu }), \qquad
D_{J,\nu }^-(\lambda )=\exp ({\bf d}_{\nu }^-(\lambda )\cdot {\bf
H}_{\nu -1}).
\end{eqnarray}
Here ${\bf d}_{\nu }^{\pm}(\lambda )=(d_{\nu ,1}^{\pm},\dots ,d_{\nu
,r}^{\pm}) $ is a vector in the root space and
\begin{eqnarray}\label{eq:3.2.9}
{\bf H}_{\eta}= \left( {2H_{\eta ,1}  \over (\alpha _{\eta,1},
\alpha _{\eta,1}) }, \dots , {2H_{\eta ,r}  \over (\alpha _{\eta,r},
\alpha_{\eta,r}) } \right), \qquad
({\bf d}_{\nu }^{\pm}(\lambda ), {\bf H}_{\eta})=
\sum_{k=1}^{r}{2d_{\nu ,k}^{\pm}(\lambda )H_{\eta ,k}  \over (\alpha
_{\eta ,k}, \alpha _{\eta ,k}) },
\end{eqnarray}
where $\alpha _{\eta ,k} $ is the $k$-th simple root of ${\frak  g} $ with
respect to the ordering $\Delta _{\eta}^+$  and $H_{\eta ,k} $ are their
dual elements in the Cartan subalgebra ${\frak  h} $.

\section{The Gauge Group Action}\label{sec:4}

Before proceeding with the study of the gauge-equivalent systems,
the following remark is in order:

We can use the gauge transformation commuting with $J $ to
simplify $Q $; in particular, we can remove all components of $Q $
in $\mathfrak{ g}_0 $; effectively, this means that our $Q(x,t)=
Q_+(x,t) + Q_{-}(x,t)\in \mathfrak{g}_+\cup \mathfrak{g}_- $ can
be viewed as a local coordinate in the co-adjoint orbit
$\mathcal{M}_J\simeq \mathfrak{g}\backslash \mathfrak{g}_0 $:
\begin{eqnarray}\label{eq:KF14.4}
Q_+(x,t) = \sum_{\alpha \in\theta _+}^{} q_\alpha(x,t) E_{\alpha
}, \qquad Q_{-}(x,t) =\sum_{\alpha \in \theta _-}^{} p_\alpha(x,t)
E_{-\alpha }.
\end{eqnarray}

\subsection{The class of the gauge equivalent NLEE's}\label{ssec:4.1}

The notion of gauge equivalence allows one to associate to any Lax pair
of the type (\ref{eq:1.5}), (\ref{eq:1.7})
an
equivalent one \cite{VYa}, solvable by the inverse scattering
method for the gauge equivalent linear problem:
\begin{eqnarray}\label{eq:2.3}
\widetilde{L}\widetilde{\psi }(x,t,\lambda )\equiv \left({\rm i}
{{\rm d} \over {\rm d}
x}-\lambda S(x,t) \right) \widetilde{\psi }(x,t,\lambda )=0, \nonumber\\
\widetilde{M}\widetilde{\psi }(x,t,\lambda )\equiv \left(i{d
\over dt} -2i\lambda \mbox{ad}_{{\cal S}}^{-1}{\cal S}_x -2\lambda
^2 {\cal S}
 \right) \widetilde{\psi }(x,t,\lambda)=0,
\end{eqnarray}
where $\widetilde{\psi }(x,t,\lambda ) = g^{-1}(x,t)\psi
(x,t,\lambda )$, $ S = \mbox{Ad}_{g}\cdot J \equiv
g^{-1}(x,t)Jg(x,t), $ and $g(x,t)=m_\nu (x,t,0) $ is FAS at
$\lambda =0 $. The functions $m_\nu (x,t,\lambda)$ are analytic
with respect to $\lambda$ in each sector $\Omega_\nu$ and do not
lose their analyticity for $\lambda=0$ (in the case of potential
on compact support). From the integral representation for the FAS
$m_\nu (x,t,\lambda)$ at $\lambda=0$, it follows that
\[
m_1(x,t,0)= \cdots = m_\nu(x,t,0)=\cdots = m_{2M}(x,t,0).
\]
Therefore, the gauge group action is well defined. The
zero-curvature condition $[\widetilde{L},\widetilde{M}]=0 $ gives:
\begin{eqnarray}\label{eq:2.5}
i {d {\cal S} \over dt } + 2 { d\over dx } \left(\mbox{ad}_{{\cal
S}}^{-1} {d {\cal S}\over dx }\right) =0.
\end{eqnarray}
Both Lax operators $L(\lambda)$ and $\widetilde{L}(\lambda)$ have
equivalent spectral properties and spectral data and therefore,
the classes of NLEE's related to them are equivalent.

Following \cite{AKNS}, one can consider more general
$\widetilde{M} $-operators of the form:
\begin{equation}\label{eq:M-op}
\widetilde{M}(\lambda)\widetilde{\Psi}\equiv {\rm
i}{d\widetilde{\Psi} \over dt}+\left(
\sum_{k=1}^{N}\widetilde{V}_k(x,t)\lambda^k
\right)\widetilde{\Psi}(x,t,\lambda)=0, \qquad f(\lambda ) =
\lim_{x\to\pm\infty } \widetilde{V}(x,t,\lambda ),
\end{equation}
where $\widetilde{V}(x,t,\lambda
)=\sum_{k=1}^{N}\widetilde{V}_k(x,t)\lambda^k$. The Lax
representation $[\widetilde{L}(\lambda),
\widetilde{M}(\lambda)]=0$ leads to  recurrent relations between
$\widetilde{V}_k(x,t)=\widetilde{V}_{k}^{\rm
f}+\widetilde{V}_{k}^{\rm d}$
\begin{eqnarray}\label{eq:RecLax}
&& \widetilde{V}_{k+1}^{\rm f}(x,t)\equiv \pi_{\cal
S}(\widetilde{V}_{k+1}) =\widetilde{\Lambda}_\pm
\widetilde{V}_k^{\rm f}(x,t)+{\rm i} \ad_{\cal S}^{-1} [C_k,\ad_{\cal S}^{-1}{\cal S}_x(x,t)], \\
&& \widetilde{V}_{k}^{\rm d}(x,t)\equiv (\openone -\pi_{\cal
S})(\widetilde{V}_{k}) =\widetilde{C}_k + \int_{\pm \infty}^x
dy\,[ \ad_{\cal S}^{-1}{\cal S}_x(y,t), \widetilde{V}_k^{\rm
f}(y,t)], \qquad k=1,...,N; \nonumber
\end{eqnarray}
where $ \pi_{\cal S}=\ad_{\cal S}^{-1}\circ\ad_{\cal S}$ and
$\widetilde{C}_k=(\openone -\pi_{\cal S})\widetilde{C}_k$ are
block-diagonal integration constants, for details see, e.g.
\cite{AKNS,ForKu*83}. These relations are resolved by the
recursion operators (\ref{eq:Lamb}):
\begin{eqnarray}\label{eq:Lamb}
\widetilde{\Lambda}=\mbox{Ad}_{g}\cdot \Lambda={1\over
2}\left(\widetilde{\Lambda}_+ + \widetilde{\Lambda}_-\right),
\qquad \widetilde{\Lambda}_\pm =\mbox{Ad}_{g} \cdot\Lambda _\pm
\end{eqnarray}
where
\begin{eqnarray}\label{eq:Lambd}
 \Lambda _\pm Z &=& \mbox{ad}_J^{-1}  (\openone -\pi_0) \left( i {dZ\over dx } +
 [q(x), Z(x)] + i \left[ q(x) , \int_{\pm\infty }^{x} dy \; \pi_0 [q(y),
 Z(y)] \right] \right),
\end{eqnarray}
and we assume that $Z \equiv \pi_{0} Z \in \mathcal{M}_{\cal S}$,
where $\pi_{0}=\mbox{ad}_J^{-1} \circ \mbox{ad}_J$ is the
projector onto the off-diagonal part. As a result, we obtain that
the class of (generically nonlocal) multi-component Heisenberg
feromagnet (MHF) type models, solvable by the ISM, have the form:
\begin{equation}\label{eq:NLEE}
i\ad_{\cal S}^{-2} {d{\cal S} \over dt } =
\sum_{k=0}^{N}\widetilde{\Lambda}_\pm^{N-k}
\left[\widetilde{C}_k,\ad_{\cal S}^{-2}{\cal S}(x,t)\right],
\qquad f(\lambda ) =\left(
\begin{array}{cc} f^+(\lambda ) & 0 \\ 0& f^-(\lambda
)\end{array}\right),
\end{equation}
where  $f(\lambda ) =\sum_{k=0}^{N}\widetilde{C}_k\lambda ^{N-k} $
determines their dispersion law. The NLEE (\ref{eq:2.5}) become
local if $f(\lambda )=f_0(\lambda ){\cal S} $, where $f_0(\lambda
) $ is a scalar function. In particular, if $f(\lambda )=-2\lambda
^2{\cal S} $ we get the MHF eqn. (\ref{eq:2.5}).

\subsection{The Minimal Set of Scattering Data for $L(\lambda)$ and $\widetilde{L}(\lambda)$}\label{ssec:4.2}

We skip the details about CBC construction which can be found in
\cite{VYa} and go to the minimal set of scattering data for the
case of complex $J $ which are defined by the sets ${\cal  F}_1 $
and ${\cal  F}_2 $ as follows:
\begin{eqnarray}\label{eq:3.2.10}
{\cal  F}_1= \mathop{\cup}\limits_{\nu =1}^{2M}{\cal  F}_{1,\nu }, \qquad
{\cal  F}_2= \mathop{\cup}\limits_{\nu =1}^{2M}{\cal  F}_{2,\nu },
\nonumber\\
{\cal  F}_{J,1,\nu }= \{\rho _{J,B,\nu ,\alpha }^{\pm }(\lambda ), \, \alpha
\in \delta _{\nu }^+, \, \lambda \in l_{\nu }\} \qquad
{\cal  F}_{J,2,\nu }= \{\tau _{J,B,\nu ,\alpha }^{\pm }(\lambda ), \, \alpha
\in \delta _{\nu }^+, \, \lambda \in l_{\nu }\},
\end{eqnarray}
where
\begin{eqnarray}\label{eq:3.2.12}
\rho _{J,B,\nu ,\alpha }^{\pm }(\lambda )=\langle S_{J,\nu }^{\pm}(\lambda
)B\hat{S}_{J,\nu }^{\pm}(\lambda ), E_{\mp \alpha }\rangle , \qquad
\tau _{J,B,\nu ,\alpha }^{\pm }(\lambda )=\langle T_{J,\nu }^{\pm}(\lambda
)B\hat{T}_{J,\nu }^{\pm}(\lambda ), E_{\mp \alpha }\rangle ,
\end{eqnarray}
with $\alpha \in \delta _{\nu }^+ $, $\lambda \in l_{\nu } $ and
$B $ is a properly chosen regular element of the Cartan subalgebra
${\frak  h} $. Without loss of generality, we can take in
(\ref{eq:3.2.12}) $B=H_{\alpha } $. Note that the functions $\rho
_{J,B,\nu ,\alpha }^{\pm}(\lambda ) $ and $\tau _{J,B,\nu ,\alpha
}^{\pm}(\lambda ) $ are continuous functions of $\lambda  $ for
$\lambda \in l_{\nu } $.

If we choose $J $ in such way that $2M=|\Delta | $-- the number of
the roots of ${\frak  g} $, then to each pair of roots $\{\alpha
,-\alpha \} $ one can relate a separate pair of rays $\{l_{\alpha
}, l_{\alpha +M}\} $, and $l_{\alpha }\neq l_{\beta } $ if $\alpha
\neq \pm \beta  $. In this case, each of the subalgebras ${\frak
g}_{\alpha } $ will be isomorphic to $sl(2) $.

In order to determine the scattering data for the gauge equivalent
equations, we need to start with the FAS for these systems:
\begin{eqnarray}\label{eq:3.1}
\widetilde{m }_\nu^{\pm}(x,\lambda
)=g^{-1}(x,t)m_\nu^{\pm}(x,\lambda )g_-,
\end{eqnarray}
where $g_-= \lim_{x \to -\infty }g (x,t) $ and due to
(\ref{eq:1.7}) and $g_-=\hat{T}(0)$.  In order to ensure that the
functions $\widetilde{\xi }^{\pm}(x,\lambda ) $ are analytic with
respect to $\lambda $, the scattering matrix $T(0) $ at $ \lambda
=0 $ must belong to ${\frak  H}\otimes {\frak  G}_0$, where
${\frak  H} $ is the corresponding Cartan subgroup and  ${\frak
G}_0 $ is the subgroup, that corresponds to the subalgebra ${\frak
g}_0$. Then, equation (\ref{eq:3.1}) provides the fundamental
analytic solutions of $\widetilde{L} $. We can calculate their
asymptotics for $x\to\pm\infty  $ and thus establish the relations
between the scattering matrices of the two systems:
\begin{eqnarray}\label{eq:3.2}
\lim_{x \to -\infty }\widetilde{\xi }^+(x,\lambda )= e^{-{\rm i}
\lambda Jx}T(0)S^+_J(\lambda )\hat{T}(0) \qquad \lim_{x \to \infty
}\widetilde{\xi }^+(x,\lambda )= e^{-{\rm i} \lambda
Jx}T^-_J(\lambda )D^+_J(\lambda )\hat{T}(0)
\end{eqnarray}
with the result: $ \widetilde{T}(\lambda )= T(\lambda
)\hat{T}(0)$. The factors in the corresponding Gauss
decompositions are related by:
\begin{eqnarray}\label{eq:3.4}
\widetilde{S}_J^{\pm}(\lambda )= T(0)S_J^{\pm}(\lambda
)\hat{T}(0), \qquad \widetilde{T}_J^{\pm}(\lambda
)=T_J^{\pm}(\lambda ) \nonumber\qquad
\widetilde{D}_J^{\pm}(\lambda )=D_J^{\pm}(\lambda )\hat{T}(0).
\end{eqnarray}
On the real axis, again, the FAS $\widetilde{\xi }^+(x,\lambda ) $
and $\widetilde{\xi }^-(x,\lambda ) $ are related by $
\widetilde{\xi }^+(x,\lambda )=\widetilde{\xi }^-(x,\lambda
)\widetilde{G}_{J,0}(\lambda ) $ with the normalization condition
$\widetilde{\xi }(x,\lambda =0)=\openone $ and
$\widetilde{G}_{J,0}(\lambda )=\widetilde{S}_J^+(\lambda
)\hat{\widetilde{S}}_J^-(\lambda ) $ again can be considered as a
minimal set of scattering data.

The minimal set of scattering data for the gauge-equivalent CBC
systems  are defined by the sets $\widetilde{\cal  F}_1 $ and
$\widetilde{\cal  F}_2 $ as follows:
\begin{eqnarray}\label{eq:3.2.10a}
\widetilde{\cal  F}_1= \mathop{\cup}\limits_{\nu
=1}^{2M}\widetilde{\cal  F}_{1,\nu }, \qquad \widetilde{\cal F}_2=
\mathop{\cup}\limits_{\nu =1}^{2M}\widetilde{\cal  F}_{2,\nu },
\nonumber\\
\widetilde{\cal  F}_{1,\nu }= \{\widetilde{\rho }_{B,\nu ,\alpha
}^{\pm }(\lambda ), \, \alpha \in \delta _{\nu }^+, \, \lambda \in
l_{\nu }\} \qquad \widetilde{\cal  F}_{2,\nu }= \{\widetilde{\tau
}_{B,\nu ,\alpha }^{\pm }(\lambda ), \, \alpha \in \delta _{\nu
}^+, \, \lambda \in l_{\nu }\},
\end{eqnarray}
where
\begin{eqnarray}\label{eq:3.2.12a}
\widetilde{\rho }_{J,B,\nu ,\alpha }^{\pm }(\lambda )=\langle
T_J(0)S_{J,\nu }^{\pm}(\lambda )B\hat{S}_{J,\nu }^{\pm}(\lambda
)\hat{T}_J(0), E_{\mp \alpha }\rangle , \quad \widetilde{\tau
}_{J,B,\nu ,\alpha }^{\pm }(\lambda )=\langle T_{J,\nu
}^{\pm}(\lambda )B\hat{T}_{J,\nu }^{\pm}(\lambda ), E_{\mp \alpha
}\rangle ,
\end{eqnarray}
with $\alpha \in \delta _{\nu }^+ $, $\lambda \in l_{\nu } $ and
$B $ is again a properly chosen regular element of the Cartan
subalgebra ${\frak  h} $. Without loss of generality, we can take
in (\ref{eq:3.2.12a}) $B=H_{\alpha } $ (as in (\ref{eq:3.2.12})).
The functions $\widetilde{\rho }_{B,\nu ,\alpha }^{\pm}(\lambda )
$ and $\widetilde{\tau }_{B,\nu ,\alpha }^{\pm}(\lambda ) $ are
continuous functions of $\lambda  $ for $\lambda \in l_{\nu }$,
and have the same analyticity properties as the functions  $\rho
_{B,\nu ,\alpha }^{\pm}(\lambda ) $ and $\tau _{B,\nu ,\alpha
}^{\pm}(\lambda ) $.

\subsection{Integrals of Motion and Hierarchies of Hamiltonian Structures}\label{ssec:4.3}

If $q(x,t) $ evolves according to the MNLS (\ref{eq:NLS-so5}),
then
 \begin{equation}\label{eq:dS_J}
 i {dS_{J}^{\pm}  \over dt } -2\lambda ^2 [J, S_{J}^{\pm} (t,\lambda )] =0,
 \qquad  i {dT_{J}^{\pm}  \over dt } -2\lambda ^2 [J, T_{J}^{\pm}
 (t,\lambda )] =0, \qquad {dD_{J}^{\pm}  \over dt }=0.
 \end{equation}
 This means that the MNLS eq. (\ref{eq:NLS-so5}) has four series of
 integrals of motion. This is due to the special (degenerate) choice of
 the dispersion law $f_{\rm MNLS}=-2\lambda ^2J $. We have to remember,
 however, that only two of these four series are in involution, which in
 turn is related to the non-commutativity of the subalgebra
 $\mathfrak{g}_J $.

Both classes of NLEE's are infinite dimensional, completely
integrable Hamiltonian systems and possess hierarchies of
Hamiltonian structures.

The phase space ${\cal  M}_{\rm MNLS} $ is the linear space of all
off-diagonal matrices $q(x,t) $ tending fast enough to zero for
$x\to\pm\infty $. The hierarchy of pair-wise compatible symplectic
structures on ${\cal  M}_{\rm MNLS} $ is provided by the $2
$-forms:
\begin{equation}\label{eq:ome-nls}
\Omega _{\rm MNLS}^{(k)} = {\rm i } \int_{-\infty }^{\infty } dx
\tr \left( \delta q(x,t) \wedge \Lambda ^k [J, \delta q(x,t) ]
\right),
\end{equation}
where $\Lambda= (\Lambda_+ + \Lambda_-)/2 $ is the generating
(recursion) operator for (\ref{eq:1.5}) defined in
(\ref{eq:Lambd}). The symplectic forms $\Omega _{\rm MNLS}^{(k)}$
can be expressed in terms of the scattering data for $L(\lambda)$:
\begin{eqnarray}\label{eq:3.2.a}
&& \Omega _{\rm MNLS}^{(k)} = {c_k \over 2\pi }\sum_{\nu =1}^{M}
\int_{\lambda \in l_{\nu }\cup l_{M+\nu } } d\lambda \lambda ^k
\left( \Omega
_{0,\nu }^+(\lambda ) - \Omega _{0,\nu }^-(\lambda )\right), \nonumber\\
&& \Omega _{0,\nu }^\pm(\lambda ) = \left\langle \hat{D}_{J,\nu }
^\pm(\lambda ) \hat{T}_{J,\nu }^\mp(\lambda ) \delta T_{J,\nu }^\mp(\lambda )
D_{J,\nu }^\pm(\lambda ) \wedge \hat{S}_{J,\nu }^\pm(\lambda ) \delta
S_{J,\nu }^\pm(\lambda ) \right\rangle .
\end{eqnarray}
Note that the kernels of $\Omega _{\rm MNLS}^{(k)}$ differ only by
the factor $\lambda^k$ so all of them can be cast into canonical
form simultaneously.

The phase space ${\cal  M}_{\rm MHF} $ of the gauge equivalent to
the MNLS systems is the manifold of all ${\cal S}(x,t) $,
satisfying appropriate boundary conditions. The family of
compatible $2 $-forms is:
\begin{equation}\label{eq:ome-hf}
\widetilde{\Omega} _{\rm MHF}^{(k)} = {i \over 4} \int_{-\infty
}^{\infty } dx \tr \left( \delta S^{(0)} \wedge
\widetilde{\Lambda} ^k [S^{(0)}, \delta S^{(0)}(x,t) ] \right).
\end{equation}
Again, like for the "canonical" MNLS models, the symplectic forms
$\Omega _{\rm MHF}^{(k)}$ for their gauge-equivalent MHF's can be
expressed in terms of the scattering data for
$\widetilde{L}(\lambda)$:
\begin{eqnarray}\label{eq:3.2.b}
&& \widetilde{\Omega} _{\rm MHF}^{(k)} = {c_k \over 2\pi
}\sum_{\nu =1}^{M} \int_{\lambda \in l_{\nu }\cup l_{M+\nu } }
d\lambda \lambda ^k \left( \widetilde{\Omega}
_{0,\nu }^+(\lambda ) - \widetilde{\Omega }_{0,\nu }^-(\lambda )\right), \nonumber\\
&& \widetilde{\Omega} _{0,\nu }^\pm(\lambda ) = \left\langle
\hat{\widetilde{D}}_{J,\nu } ^\pm(\lambda )
\hat{\widetilde{T}}_{J,\nu }^\mp(\lambda ) \delta
\widetilde{T}_{J,\nu }^\mp(\lambda ) \widetilde{D}_{J,\nu
}^\pm(\lambda ) \wedge \hat{\widetilde{S}}_{J,\nu }^\pm(\lambda )
\delta \widetilde{S}_{J,\nu }^\pm(\lambda ) \right\rangle .
\end{eqnarray}
The spectral theory of these two operators $\Lambda$ and
$\widetilde{\Lambda }$ underlie all the fundamental properties of
these two classes of gauge equivalent NLEE, for details see
\cite{VYa}.  Note that the gauge transformation relates, in a
nontrivial manner, the symplectic structures, i.e. $\Omega _{\rm
MNLS}^{(k)} \simeq\widetilde{\Omega} _{\rm MHF}^{(k+2)} $
\cite{RST,VYa}.

\section{Example ${\frak g}\simeq so(5,\bbbc)$}\label{sec:5}

Here, we  consider a MNLS model with the Lax operators
$L(\lambda)$ and $M(\lambda)$ belonging to $so(5,\bbbc)$ Lie algebra.

\begin{figure}\label{fig:1}

\caption{The continuous spectrum of the Lax operator $L(\lambda )
$ related to (\ref{eq:6MNLS}).}
\begin{center}
\special{em:linewidth 0.4pt}
\unitlength 1mm
\linethickness{0.4pt}
\begin{picture}(95.11,94.77)
\emline{10.00}{30.00}{1}{90.00}{70.00}{2}
\emline{10.00}{70.00}{3}{90.00}{30.00}{4}
\emline{10.00}{50.00}{5}{90.00}{50.00}{6}
\emline{50.00}{7.00}{7}{50.00}{90.00}{8}
\emline{50.00}{90.00}{9}{50.00}{90.00}{10}
\put(80.00,50.00){\vector(1,0){0.2}}
\emline{20.00}{50.00}{11}{80.00}{50.00}{12}
\put(50.00,80.00){\vector(0,1){0.2}}
\emline{50.00}{20.00}{13}{50.00}{80.00}{14}
\put(80.33,57.33){\makebox(0,0)[cc]{$\Omega_1$}}
\put(65.00,71.00){\makebox(0,0)[cc]{$\Omega_2$}}
\put(34.67,71.00){\makebox(0,0)[cc]{$\Omega_3$}}
\put(19.00,57.33){\makebox(0,0)[cc]{$\Omega_4$}}
\put(19.33,39.00){\makebox(0,0)[cc]{$\Omega_5$}}
\put(34.67,26.33){\makebox(0,0)[cc]{$\Omega_6$}}
\put(64.67,26.33){\makebox(0,0)[cc]{$\Omega_7$}}
\put(80.33,39.00){\makebox(0,0)[cc]{$\Omega_8$}}
\put(97.33,49.67){\makebox(0,0)[cc]{$l_1$ (${\frak g}_0$)}}
\put(92.00,69.33){\makebox(0,0)[cc]{$l_2$}}
\put(92.33,26.00){\makebox(0,0)[cc]{$l_8$}}
\put(51.00,4.00){\makebox(0,0)[cc]{$l_7$}}
\put(51.67,91.00){\makebox(0,0)[cc]{$l_3$}}
\put(9.00,70.00){\makebox(0,0)[cc]{$l_4$}}
\put(6.67,49.67){\makebox(0,0)[cc]{$l_5$}}
\put(8.33,27.00){\makebox(0,0)[cc]{$l_6$}}
\put(90.33,90.00){\circle{6.55}}
\put(90.33,90.00){\makebox(0,0)[cc]{$\lambda$}}
\end{picture}
\end{center}
\end{figure}
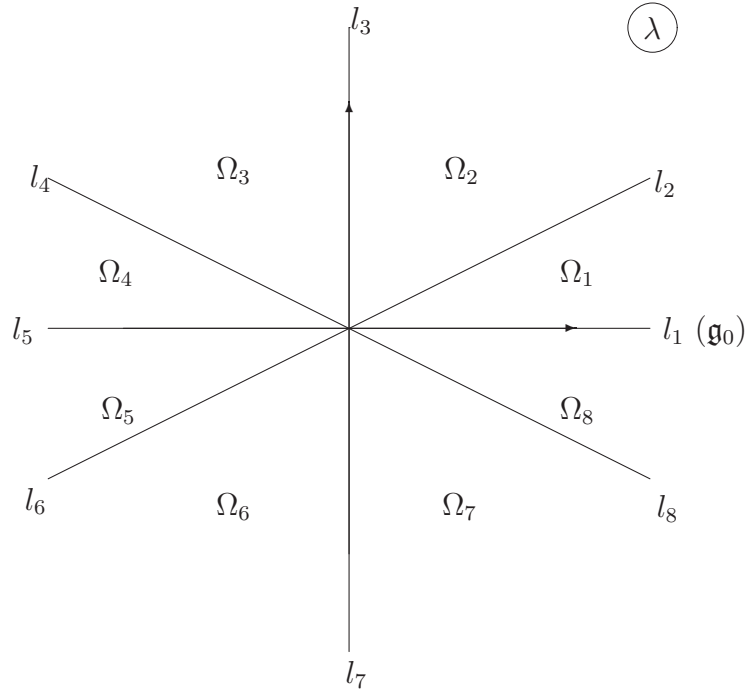

This algebra has 4 positive roots: $\alpha_1=e_1-e_2$,
$\alpha_2=e_2$, $\alpha_3=\alpha_1+\alpha_2$ and
$\alpha_4=\alpha_1+2\alpha_2$. Let us choose also $J $ to be a
degenerate Cartan element ($\alpha _1(J)=0 $) so that the set of
roots $\Delta _1^+=\{ \alpha _2,\alpha_3,\alpha _4\}$ of $so(5)$,
for which $\alpha (J)\neq 0 $ labels the coefficients of the
potential $q(x,t)$:
\begin{eqnarray}\label{grah:ex-nls}
q(x,t)&\equiv & \sum_{\alpha \in \Delta _1^+}^{} (q_{\alpha
}E_{\alpha } + p_{\alpha }E_{-\alpha }) =
\left(\begin{array}{ccccc} 0 & 0 & q_{11} & q_{12} & 0\\
0 & 0 & q_1 & 0 & q_{12} \\ p_{11} & p_1 & 0 & q_1 & -q_{11}\\
p_{12} & 0 & p_1 & 0 & 0\\  0 & p_{12} & -p_{11} & 0 & 0\\
\end{array} \right); \qquad
J=\mbox{diag}\,(a,a,0,-a,-a)\nonumber.
\end{eqnarray}
Here, $q_1$ and $p_1$ are related to the root $\alpha_2$; the
labels $mn$ in $q_{mn}(x,t)$ and $p_{mn}(x,t)$ refer to the roots
$(mn)\leftrightarrow m\alpha_1+n\alpha_2$. The continuous spectrum of the Lax operator $L(\lambda)$ related to (\ref{eq:6MNLS}) is depicted  on fig.1. Then, the corresponding
MNLS type system is of the form:
\begin{eqnarray}\label{eq:6MNLS}
\hspace*{-7mm} i{\partial q_{12}\over \partial t}+ {1\over
2a}{\partial ^2q_{12}\over \partial x^2}+{1 \over
a}q_{12}(q_1p_1+q_{11}p_{11}+q_{12}p_{12}) + {i\over
a}q_{1}q_{11,x}-{i\over a}q_{11}q_{1,x}&=&0, \nonumber\\
\hspace*{-7mm} i{\partial q_{11}\over \partial t}+ {1\over
a}{\partial ^2q_{11}\over \partial x^2}+{1 \over
a}q_{11}(q_1p_1+q_{11}p_{11}+{1\over 2}q_{12}p_{12}) + {i\over
a}q_{12}p_{1,x}+{i\over 2a}q_{12,x}p_{1}&=&0, \nonumber\\
\hspace*{-7mm}  i{\partial q_{1}\over \partial t}+ {1\over
a}{\partial ^2q_{1}\over \partial x^2}+{1 \over
a}q_{1}(q_1p_1+q_{11}p_{11}+{1\over 2}q_{12}p_{12}) - {i\over
a}q_{12}p_{11,x}-{i\over 2a}q_{12,x}p_{11}&=&0, \\
\hspace*{-7mm} i{\partial p_{1}\over \partial t}- {1\over
a}{\partial ^2p_{1}\over \partial x^2}-{1 \over
a}p_{1}(q_1p_1+q_{11}p_{11}+{1\over 2}q_{12}p_{12}) - {i\over
a}p_{12}q_{11,x}-{i\over 2a}p_{12,x}q_{11}&=&0, \nonumber\\
\hspace*{-7mm} i{\partial p_{11}\over \partial t} -{1\over
a}{\partial ^2p_{11}\over \partial x^2}-{1 \over
a}p_{11}(q_1p_1+q_{11}p_{11}+{1\over 2}q_{12}p_{12}) + {i\over
a}p_{12}q_{1,x}+ {i\over 2a}p_{12,x}q_{1}&=&0, \nonumber\\
\hspace*{-7mm} i{\partial p_{12}\over \partial t}- {1\over
2a}{\partial ^2p_{12}\over \partial x^2}-{1 \over
a}p_{12}(q_1p_1+q_{11}p_{11}+q_{12}p_{12}) + {i\over
a}p_{1}p_{11,x}-{i\over a}p_{11}p_{1,x}&=&0. \nonumber
\end{eqnarray}
In order to evaluate the gauge equivalent recursion operator,
 we  use  the gauge covariant approach. First, we  need
 to express any function $f(K)$ of the the operator
 $K=\mbox{ad}_J$ through the projectors onto its eigensubspaces.
Our choice of $J $ means that $K=\ad_J $ has five different
eigenvalues:
 $-2a $, $-a $, $0 $, $a $ and $2a $. Then, the  minimal characteristic
 polynomial for $K $ is  $ K(K^2-a^2)(K^2-4a^2) =0$.
 Let us also introduce the  projectors onto the eigensubspaces of $K $ as
 follows:
 \begin{eqnarray*}\label{eq:1_5)}
 \pi_{\pm 2} = {K(K^2-a^2) (K\pm 2a)\over 24a^4 }, \quad
 \pi_{\pm 1} = -{K(K^2-4a^2) (K\pm a)\over 6a^4 }, \quad
 \pi_0 =  {(K^2-a^2) (K^2-4a^2)  \over 4a^4 }.
 \end{eqnarray*}
 Using the characteristic equation $ K(K^2-a^2)(K^2-4a^2) =0$, it is easy to
 check that $\pi_{j} $ are orthogonal projectors; i.e. they
 satisfy: $ \pi_{j} \pi_{k} = \delta _{jk}\pi_{j}$ for all $j,k=\pm 2 $,
 $\pm 1 $, $0 $ and that
 $  K\pi_{\pm 2} = \pm 2a \pi_{\pm 2}$,
 $K\pi_{\pm 1} = \pm a \pi_{\pm 1}$, $K\pi_0 = 0$.
 Thus any function $f(K) $ can be expressed in terms of these projectors:
 $
 f(K) = f(2a)\pi_{2} + f(a)\pi_{1} + f(0)\pi_{0}+ f(-a)\pi_{-1}
 f(-2a)\pi_{-2}
 $
 provided $f(\lambda ) $ is regular for $\lambda =\pm 2a $, $\pm a $
 and $0 $. Note also that $\ad_J=K
 $ introduces a grading on $\mathfrak{g} =\mathop{\oplus}\limits_{j=-2}^2
 \mathfrak{g}_j $ and the projectors $\pi_j $ project precisely onto
 $\mathfrak{g}_j $.  Obviously, $\mathfrak{g}_j=\pi_j \mathfrak{g} $,
 $\mathfrak{g}_0\equiv \mathfrak{g}_J $ and $\mathcal{M}_J \simeq
 \mathfrak{g}\backslash \mathfrak{g}_J $.

Then, applying a gauge transformation, one can  recalculate easily
the projectors on the eigensubspaces of
 $\ad_{S(x)}\equiv \widetilde{K}(x) = g_0^{-1}Kg_0(x,t) $:
 \begin{eqnarray*}\label{eq:1_5}
 \widetilde{\pi}_{\pm 2} = {\widetilde{K}(\widetilde{K}^2-a^2) (
 \widetilde{K}\pm 2a)\over 24a^4 }, \quad \widetilde{\pi}_{\pm 1} =
 -{\widetilde{K}(\widetilde{K}^2-4a^2) (\widetilde{K}\pm a)\over 6a^4 },
 \quad \widetilde{\pi}_0 = {(\widetilde{K}^2-a^2) (\widetilde{K}^2-4a^2)
 \over 4a^4 }.
 \end{eqnarray*}
Using these formulae, one can cast also, the gauge-equivalent
MHF-type system  (\ref{eq:2.5}) in the form:
 \begin{equation}\label{eq:2.5a}
 iS_t - {5  \over 4a^2 } [ S, S_{xx}] + {1  \over 4a^4 }
 \left( (\ad_S)^3 S_x \right)_x =0,
 \end{equation}
 where ${\cal S}$ is constrained by ${\cal S}({\cal S}^2-a^2)^2=0$.
 In addition, the operator $\widetilde{K}(x,t) $ satisfies the equation
 $ \widetilde{K}(\widetilde{K}^2-a^2)(\widetilde{K}^2-4a^2) =0$.

Now, let us apply to the system (\ref{eq:6MNLS}) the following
reduction: $L(\lambda)=-L(\lambda^*)^\dag$. This implies, that the
potential matrix $Q(x,t)$ is hermitian, i.e. that $p_\alpha
=q_\alpha^*$, and, that the matrix elements of the Cartan element
$J$ are real. As a result, we get the following 3-component MNLS
system for the complex-valued fields $q_{1}(x,t)$, $q_{11}(x,t)$
and $q_{12}(x,t)$:
 \begin{eqnarray}
 &&i{dq_{12}\over dt}+ {1\over 2a}{d^2q_{12}\over dx^2}-{1 \over a}q_{12}
 (|q_1|^2 +|q_{11}|^2 + |q_{12}|^2) + {i\over a} q_{1}q_{11,x}-
 {i\over a}q_{11}q_{1,x}=0 \nonumber\\
 &&i{dq_{11}\over dt}+ {1\over a}{d^2q_{11}\over dx^2}-{1 \over a}
 q_{11}(|q_1|^2 + |q_{11}|^2 + {1\over 2} |q_{12}|^2 ) + {i\over a}
 q_{12} q^*_{1,x}+ {i\over 2a}q_{12,x}q^*_{1}=0 \\
 \label{eq:NLS-so5}
 &&i{dq_{1}\over dt}+ {1\over a}{d^2q_{1}\over dx^2}-{1 \over a}
 q_{1} (|q_1|^2 + |q_{11}|^2 + {1\over 2} |q_{12}|^2 ) - {i\over a}
 q_{12}q^*_{11,x} - {i\over 2a} q_{12,x}q^*_{11}=0,\nonumber
 \end{eqnarray}
Using the well-known isomorphism between the algebras
$so(5,\bbbc)$ and $sp(4,\bbbc)$ \cite{Helg}, due to the purely
Lie-algebraic nature of the Lax representation, one can convert
this 3-component MNLS system in the typical representation of
$sp(4,\bbbc)$\footnote{this representation is equivalent  to the
spinor representation of $so(5,\bbbc)$ \cite{GoGr,Helg}}. This
system is equivalent to the 3-component one, describing $F=1$
spinor Bose-Einstein condensate \cite{IMW04,kagg} in one
dimensional approximation.

Finally, applying the reduction
$\widetilde{L}(\lambda)=-\widetilde{L}(\lambda^*)^\dag$ we obtain
that the reduced model, that corresponds to (\ref{eq:2.5a}) will
be constrained by the condition, that the matrix ${\cal S}$ must
be hermitian: ${\cal S}={\cal S}^\dag$.

\section{Conclusions}\label{sec:6}

We will finish this article with several concluding remarks.
 In order to obtain the soliton solutions for the gauge equivalent MHF
 systems, one needs to apply the Zakharov--Shabat dressing method to a regular
 FAS $\widetilde{\chi}_{(0)}^{\pm}(x,\lambda ) $ of $\widetilde{L} $ with potential
 ${\cal S}_{(0)} $. Thus, one gets a new singular solution   $\widetilde{\chi
 }_{(1)}^{\pm}(x, \lambda ) $ of the Riemann--Hilbert problem  with
 singularities located at prescribed positions $\lambda _1^{\pm} $.  It is
 related to the regular one by the dressing factors
 $\widetilde{u}(x,\lambda)$.
 The dressing method for the generalised Zakharov-Shabat systems (related
 to semi-simple Lie algebras) is developed in \cite{Za*Mi,G*86} , \cite{VSG*87}, \cite{HSAS} and \cite{Ivanov}.

 To
MNLS systems and their gauge equivalent MHF ones, one can apply
the analysis \cite{VYa} and derive the completeness relations for
the corresponding system of "squared" solutions. Such analysis
will allow one to prove the pair-wise compatibility of the
Hamiltonian structures and eventually, to derive their
action-angle variables, see e.g. \cite{ZM} and \cite{BS} for the
${\bf A}_r$-series.

The approach presented here allows one to consider CBC systems
with more general    $\lambda$- dependence, like the Principal
Chiral field models and other relativistic invariant field
theories \cite{Za*Mi}.

Finally, some open problems are: 1) to study the internal
structure of the soliton solutions and soliton interactions (for
both types of systems); 2) to study reductions of the gauge
equivalent MHF systems and the spectral decompositions for the
relevant recursion operator.

\subsection*{Acknowledgements}

The author has the pleasure to thank professors E. V. Ferapontov, V. S. Gerdjikov, R. I. Ivanov,
N, A, Kostov and A. V. Mikhailov  for numerous stimulating discussions. This material is based upon works supported by the Science Foundation of
Ireland (SFI), under Grant No. 09/RFP/MTH2144.

\end{document}